\documentclass{epl}

\title{Time dependent correlations of inflationary perturbations}
\author{W.-L. Lee\inst{1} \and L.-Z. Fang\inst{2}}
\institute{
  \inst{1} Institute of Physics, Academia Sinica, Taipei, Taiwan 11529,
Republic of China\\
  \inst{2} Department of Physics, University of Arizona, Tucson,
AZ 85721, U.S.A.
}

\begin{document}

\maketitle

\begin{abstract}

We show that if the primordial classical perturbations were 
generated by the gravitational particle creation during inflation,
and followed by an evolution of quantum-to-classical transition, 
the time dependent correlation of these perturbations is 
long-tailed with a correlation time larger than the Hubble-time. 
Consequently, the inflationary perturbations are locally scale-scale 
correlated. Hence, the interaction of the fields during inflation 
can be explored via the detection of the local scale-scale correlation 
of the CMB fluctuations.

\end{abstract}

It is fundamentally important to understand the nature of the primordial
cosmological perturbations which are responsible for the formation of 
structures that we see today. The inflation paradigm for the early universe
assumes that the cosmic perturbations are initiated by the process of
particle creation from vacuum in the background gravitational field of
the expanding universe\cite{haw}. The subsequent decoherence of the
quantum fluctuations leads to a quantum-to-classical transition\cite{sak},
and gives rise to the initial perturbations in the radiation dominated era
on scales beyond the Hubble radius. Therefore, the earliest evolution of
the inflationary cosmological fluctuations is governed by the
gravitational particle creation accompanying the quantum-to-classical
transition. Recently, the mechanism of the gravitational particle creation
is re-emphasized with the development of the quintessential inflation
model\cite{pee}. However, it still seems to lack some testable predictions
of these mechanisms.

In this {\it Letter}, we will show that the time-dependent correlation of
the primordial perturbations is one of such testable predictions. In the
``standard" inflation, i.e. the slow-roll inflation caused by a single
scalar inflaton $\phi$, the correlation time of the primordial
perturbations is longer than the Hubble time $H^{-1}$. This correlation
will entail a non-trivial observable feature -- the local scale-scale
correlation of the initial perturbations.
 
The time-dependence of cosmological perturbations is generally not
observable in non-inflationary models, because the observed mass field
does not contain two or multi-time information of the perturbations. In
contrast, the inflationary scenario provides an excellent prospect to
observing the time-dependence of the initial perturbations in a given
region on the scale of horizon. Owing to the so-called ``first out -- last
in'' feature, a fluctuation crossing over the Hubble radius $H^{-1}$ at
the moment $t^{\times}$ will yield a perturbation at the end of the
inflation, $t_f$, with the physical wavenumber given by \cite{lyth}
\begin{equation} k_p \propto e^{H(t^{\times}-t_f)}.
\end{equation} 
Consequently, the longer the physical length of the perturbations, the
earlier the time of becoming the super-horizon scaled, and vice versa.
Therefore, if perturbations with the same physical scale at two different
instants $t_1^{\times}$ and $t_2^{\times}$ are correlated during the
super-horizon evolution, the perturbations of the corresponding physical
scales $k_p(t_1^{\times})$ and $k_p(t_2^{\times})$ at the same time $t_f$
will be correlated.  Equation (1) prescribes exactly the mapping of the
correlation between two perturbations on the same physical scale with
different horizon-crossing moments to that of two perturbations at the
same time with different physical scales. The latter is actually
observable. If the length scales of perturbations are larger than the
Hubble radius $H^{-1}$ at the decoupling era, the perturbations were less
contaminated by the post inflationary reheating and other small scale
processes. Thus, one can expect that the two-scale correlation of the
cosmic background temperature fluctuations retain the information of the
time dependent behavior of the initial perturbations.

The time-dependent correlation is not a new issue. It is generally believed 
that the perturbations originated from the quantum fluctuations of vacua
during inflation have large correlation time since they vary slowly in the 
super-horizon regime. However, the rough idea of the lengthy correlation time
is inadequate to establish the detectability of the time-dependent correlation
, and one must study the temporal correlation of the perturbations in both 
quantum and classical regime to reach the equation suitable of calculating
the time correlation in practical.
 
Let us calculate the time-dependent behavior for the ``standard''
slow-roll inflation governed by a real massless scalar field $\phi$
described by the action
\begin{equation}
S=\frac{1}{2}\int d^4x\sqrt{-g}
  [\partial^{\mu}\phi\partial_{\mu}\phi -V(\phi)],
\end{equation}
where $V(\phi)$ is a self-interaction potential. Under the slow-roll
condition, the time-scale of the self-interaction is much longer than
$1/H$. That is, the self-interaction is negligible if we study
only the evolution on time scales smaller than the duration of inflation.
Without interaction, the evolution of perturbations of the free $\phi$
field mode will be coherent. Thus, the $\phi$ field self-interaction
is ineffective of violating the temporal coherence during inflation.

For the simple model Eq. (2), the dominant interaction during inflation is
the coupling between the $\phi$ field and the gravitation of the expanding
universe, governed by the gravitational scalar density $\sqrt{-g}$ in the
integral. Therefore, in the ``standard" inflation model, the origin and
evolution of the temporal coherence of the primordial perturbations are
actually determined by the dynamics of the gravitationally driven particle
creation.

In a de Sitter space, the scalar field $\phi({\bf r}, t)$ can 
be described as a superposition of free modes with coordinate and 
conjugate momentum variables given by
$\phi({\bf k}) = 
(2\pi)^{-3/2} \int \phi({\bf r})e^{-i{\bf k \cdot r}}d^3{\bf r}$
and $\pi({\bf k})=a^2\phi'({\bf k})$, respectively, where ${\bf k}$ is 
the comoving  
wavevector and $a=e^{Ht}$ represents the cosmic scale factor\cite{rey}. The
notation $\prime$ denotes the derivative with respect to the conformal
time $\tau=\int dt/a(t)=-(1/H)e^{-Ht}$. Therefore, the time-dependent
behavior of the $\phi$ field can be studied via the modes ${\bf k}$.

The quantum nature of the $\phi$ field is specified by the equal-time
commutation relation given by 
$[\phi({\bf k},\tau), \pi^{\dagger}({\bf k}, \tau)]=
  i\delta^{(3)}({\bf k}-{\bf k'})$, which yields
\begin{eqnarray}
\phi({\bf k}, \tau) & = &  \sigma_{\bf k}(\tau)a_{\bf k}+ 
   \sigma^{*}_{\bf k}(\tau)a^{\dagger}_{\bf -k}, \\ \nonumber
\pi({\bf k}, \tau)  & = &  a^2\sigma'_{\bf k}(\tau)a_{\bf k}+
   a^2\sigma^{*\prime}_{\bf k}(\tau)a^{\dagger}_{\bf -k}.
\end{eqnarray}
The $a_{\bf k}$ and $a^{\dagger}_{\bf k}$ are respectively, the 
annihilation and creation operators, satisfying commutation relations
$[a_{\bf k},a^{\dagger}_{\bf k'}]=\delta^{(3)}({\bf k}-{\bf k'})$. The
time dependence of the mode ${\bf k}$ is given by
\begin{equation}
\sigma_{\bf k}(\tau) = \frac{1}{\sqrt{2k}}
\left [H\tau - i\frac{H}{k}\right ] e^{-ik\tau},
\end{equation}

If the system is in the vacuum state at a given time $\tau_0$, we have
\begin{equation}
a_{\bf k}|0,\tau_0 \rangle =0
\end{equation}
for all ${\bf k}$. In the Schr\"odinger representation, the evolution of
the system leads the vacuum state at $\tau_0$ developing into 
$|0,\tau\rangle = S|0,\tau_0\rangle$ at a later time $\tau$, where $S$ is 
the $S$-matrix.  
The state $|0,\tau \rangle$ is no longer a vacuum state. The average
number of the particles gravitationally created in mode ${\bf k}$ is
$N_{\bf k}(\tau)= \langle 0,\tau_0| S^{-1}a^{\dagger}_{\bf k}a_{\bf -k}S
|0,\tau_0 \rangle$. 

One cannot directly calculate the time-dependent correlation of classical
density perturbations by the state $|0,\tau \rangle$, because at a different 
$\tau$ the states are quantum coherent. We should first find the
condition of quantum decoherence for the system at different times. In
the coordinate representation $\phi({\bf k})$, the Schr\"odinger wave
function of the state $|0,\tau\rangle$ is given by the product of the wave
functions for all ${\bf k}$ modes\cite{lps}:
\begin{equation}
\Psi[\phi({\bf k}), \tau] = 
\frac{1}{\sqrt{\pi}|\sigma_{\bf k}(\tau)|}
\exp \left \{ - \frac{|\phi({\bf k})|^2}{2|\sigma_{\bf k}(\tau)|^2}
  [1-i2F({\bf k}, \tau)]\right \},
\end{equation}
where 
\begin{equation}
F({\bf k}, \tau) =\frac{1}{2k\tau}=-\frac{H}{2k} e^{Ht}. 
\end{equation}
Equation (6) is actually the ground state wave function of a harmonic 
oscillator with coordinate $|\phi({\bf k})|$ which possesses a
time-dependent variance in itself.  

Thus, the quantum coherence between the coordinates at different times can
be calculated as
\begin{eqnarray}
\lefteqn{ \langle \Psi(\tau_2)|\phi^{\dagger}({\bf k})\phi({\bf k})| 
  \Psi(\tau_1) \rangle   = 
  \frac{1}{2\sqrt{\pi}|\sigma_{\bf k}(\tau_1)||\sigma_{\bf k}(\tau_2)|}
} \\ \nonumber
 & & \ \ \left \{\frac{1}{2|\sigma_{\bf k}(\tau_1)|^2}[1-i2F({\bf k}, \tau_1)]
+\frac{1}{2|\sigma_{\bf k}(\tau_2)|^2}[1+i2F({\bf k}, \tau_2)] 
  \right \}^{-3/2}.
\end{eqnarray}
When $t \gg H^{-1}$ or $|H\tau|= e^{-Ht} \ll 1$, the term $H\tau$ in
Eq. (4) can be ignored, and $|\sigma_{\bf k}(\tau)|$ becomes
$\tau$-independent.  Accordingly, 
$|\sigma_{\bf k}(\tau_1)|\simeq|\sigma_{\bf k}(\tau_2)|$, and the ratio of
the two-time matrix element to the diagonal element reduces to
\begin{equation}
\frac{|\langle \Psi(\tau_2)|\phi^*({\bf k})\phi({\bf k})|
     \Psi(\tau_1)\rangle|^2}
{|\langle \Psi(\tau_1)|\phi^*({\bf k})\phi({\bf k})|\Psi(\tau_1)\rangle|^2} 
\simeq \left\{ 1+\frac{1}{4}\left(\frac{H}{k}\right)^2\left[e^{Ht_1} - 
e^{Ht_2} \right]^2 \right\}^{-3/2}.
\end{equation}
Evidently, Eq. (9) is quickly approaching zero which implies that the
decoherence of the system for all $|t_1-t_2| \geq 1/H$, once $ke^{-Ht_1}$
and $ke^{-Ht_2}$ are less than $H$.  Thus, the relationship between the
${\bf k}$-mode fluctuations at times $t_1$ and $t_2$ can be treated as
classical perturbations provided that both physical wavelengths,
$e^{Ht_1}/k$ and $e^{Ht_2}/k$, are super-Hubble sized with the time
separation $|t_1-t_2|$ being greater than one Hubble time.

The quantum decoherence does not exclude the classical time correlation of 
perturbations\cite{hal}. The classical correlation can be derived by means 
of the Wigner distribution functions of the system. For the ${\bf k}$ mode, 
it reads
\begin{eqnarray}
W(\phi({\bf k}),\pi({\bf k}), \tau) & = & 
\int d\left(\frac{\varphi}{2\pi}\right) e^{-i\pi \varphi}
 \Psi^*\left( \phi({\bf k})-\frac{\varphi}{2}, \tau \right )
 \Psi\left( \phi({\bf k})+ \frac{\varphi}{2}, \tau \right )  \nonumber \\ 
 & = & N_{\bf k} \frac{|\sigma_{\bf k}|^2}{\pi}
  \exp\left [-\frac{|\phi({\bf k})|^2}{|\sigma_{\bf k}(\tau)|^2}\right ]
  \exp \left[-|\sigma_{\bf k}(\tau)|^2 \left |\pi({\bf k})-
   \frac{F({\bf k},\tau)}{|\sigma_{\bf k}(\tau)|^2} \phi({\bf k})\right |^2 
 \right ]. 
\end{eqnarray}
Since $|\sigma_{\bf k}(\tau)|$ is $\tau$-independent when 
$t\gg 1/H$, the first exponent on the right hand side of Eq.(10) represents 
the Gaussian probability distribution of the fluctuation of the scalar
field $\phi({\bf k})$.

The time-dependent behavior of the classical perturbations can be
extracted through the second Gaussian probability distribution in
Eq. (10). In the classical limit $\hbar\longrightarrow 0$, it yields
\begin{equation}
W(\phi({\bf k}),\pi({\bf k}), \tau) \simeq  N_{\bf k} 
\exp\left [-\frac{|\phi({\bf k})|^2}{|\sigma_{\bf k}(\tau)|^2}\right]
\delta\left(\pi({\bf k})-
\frac{F({\bf k},\tau)}{|\sigma_{\bf k}(\tau)|^2} \phi({\bf k})\right). 
\end{equation}
This shows that the classical trajectory of the ${\bf k}$ mode in phase
space is given by
\begin{equation} 
\pi({\bf k})-
   \frac{F({\bf k},\tau)}{|\sigma_{\bf k}(\tau)|^2} \phi({\bf k})=0.
\end{equation}
The degree, or the effectiveness, of classical correlation can be measured by
the relative sharpness of the classical trajectory in phase space, which
is defined as the ratio of the dispersion in momentum to the
magnitude of the average of the momentum $|\pi({\bf k})|$\cite{mo}.  
By virtue of Eq. (10), this ratio is 
\begin{equation}
\frac{1/|\sigma_{\bf k}(\tau)|}
{|(F({\bf k},\tau)/|\sigma_{\bf k}(\tau)|^2)\phi({\bf k})|}
\simeq \frac{1}{|F({\bf k}, \tau)|} \ll 1, \ \ \ {\rm if} \
  |k\tau|= (k/H)\exp(-Ht) \ll 1. 
\end{equation}
Therefore, the classical perturbations of super-Hubble size, i.e. 
$k e^{-Ht} \ll H $, are perfectly coherent. Namely, after the 
quantum-to-classical transition, the evolution of the initial 
perturbations can be calculated by the classical trajectory Eq. (12). 

The scale $t_c$ of the time-dependent correlation between classical 
perturbations of ${\bf k}$ mode can be characterized by 
\begin{equation}
\frac{1}{t_c} \simeq 
-\frac{\partial\phi({\bf k})/\partial t}{\phi({\bf k})}  =
- 
\frac{\partial \tau}{\partial t}\frac{\pi({\bf k})}{a^2 \phi({\bf k})} 
  \simeq  \frac{k^2 e^{-2Ht}}{H}.
\end{equation}
Hence, for super-Hubble sized perturbations, we have
\begin{equation}
t_cH \gg 1.
\end{equation}
That is, the correlation time of the primeval perturbations is much longer
than the Hubble time $H^{-1}$. 

This result is anticipated because the only interaction of the inflation 
in the ``standard'' model during inflation is the gravitational particle
creation. After the quantum-to-classical transition, the perturbed 
scalar field consists of classical free motion. Since there is no 
coupling between either comoving modes ${\bf k}$ or physical modes
${\bf k_{\it p}}$, the evolution of the classical perturbations of
the $\phi$ field are coherent. Therefore, one may conclude that the large 
correlation time of the initial perturbations probably is a generic 
feature for models with free inflaton, i.e., besides the gravitational 
coupling of the inflaton, there are no interactions and self-interaction 
during the slow-roll phase. This point can be seen more clearly if
we establish an equation for the classical perturbations of the $\phi$
field.

Actually, one can regard Eq.(14) as an evolution equation of the ensemble
averaged $\phi({\bf k})$, i.e. $\partial\phi({\bf k})/\partial t\simeq
-(k^2 e^{-2Ht}/H)\phi({\bf k})$. The uncertainty of this equation
can be estimated by the time scale given by
\begin{equation}
\frac{1}{t'_c} \simeq 
\left|\frac{\Delta \partial \phi({\bf k})/\partial t}{\phi({\bf k})}
\right|=\frac{\left|\Delta \pi({\bf k})\right|}{a^3\left|\phi({\bf k})
\right|} \simeq \frac{k^3e^{-3Ht}}{H^2}.
\end{equation}
We have $t'_c \gg t_c$ and therefore, the classical evolution equation
for the ${\bf k}$ mode perturbations in phase space is reasonable in
average.   

The two equations (14) and (16) of the ensemble averaged mode
$\phi({\bf k})$ can be combined into the following Langevin-like equation 
for stochastic variable $\phi({\bf k})$ 
\begin{equation}
\frac{\partial\phi({\bf k})}{\partial t} 
\simeq - \frac{k^2 e^{-2Ht}}{H}\phi({\bf k}) + \eta_{\bf k},
\end{equation}
where the small Gaussian noise $\eta_{\bf k}$ has zero mean with variance
$\simeq k^{3/2}e^{-3Ht}/H$. Therefore, the ensemble average of (17) 
yields Eq.(14), and the variance gives (16). Since Eq.(17) is linear to
$\phi({\bf k}, t)$, it is straightforward to get the equation of 
$\phi({\bf r}, t)$ in the comoving ${\bf r}$-representation. After employing
the physical coordinates in which $dx_i=e^{Ht}dr_i$ ($i=1,2,3$), Eq.(17) 
becomes
\begin{equation}
\frac {\partial \phi({\bf x}, t)}{\partial t} \simeq
  \frac{1}{H}\nabla^2_{\bf x} \phi({\bf x}, t) + \eta({\bf x},t),
\end{equation}
where $\nabla^2_{\bf x}$ stands for the Laplacian of the physical coordinates
${\bf x}$. Because the variance of the stochastic noise is scaled as 
$\langle |\eta({\bf x},t)|^2 \rangle \propto e^{-3Ht}$, 
the noise term is strongly suppressed as $Ht \gg 1$. 

Equation (18) shows that the correlation time for a mode with the physical 
wavevector ${\bf k}_p$, i.e. $\phi({\bf x}, t) \sim e^{-i{\bf k}_p\cdot {\bf x}}$,
is proportional to $k^{-2}_p$. This is typical to the diffusion-like 
soft modes in the theory for non-equilibrium systems\cite{kirk}. 
If a perturbed field consists of the superposition of such soft long 
wavelength perturbations described by Eq.(18), the time dependent 
correlation function of the perturbed $\phi$ field possesses a 
long-tail\cite{ert}. This behavior is more transparent when considering modes
characterized by $e^{-i\omega t-i{\bf k}_p\cdot {\bf x}}$. For these modes
Eq.(18) gives rise to the dispersion relation as $\omega = -ik_p^2/H$. 
Apparently, the relaxation time 
$-1/\Im\omega_{\bf k_{\it p}} \rightarrow \infty$ 
when $k_p \rightarrow 0$, and these modes lead to long-range correlations 
in general. 

Since we are only interested in the perturbations consisting of modes 
with physical wavelength larger than $H^{-1}$, the field $\phi({\bf x},t)$ 
crossing over the horizon can be prescribed by
\begin{equation}
\phi({\bf x}, t) = \int_{k_p < H}
d^{3}{\bf k}_p \hat{\phi}_{{\bf k}_p} e^{-i\omega t-i{\bf k}_p\cdot {\bf x}},
\end{equation}
where $\hat{\phi}_{{\bf k}_p}$ is the amplitude of the perturbation for modes
${\bf k}_p$. Thus, the time-dependent correlation function is determined 
by
\begin{equation}
\langle \phi({\bf x}, t_1)\phi^{*}({\bf x}, t_2) \rangle =  
 \int_{k_p < H} d^3{\bf k}_p 
\langle \hat{\phi}_{{\bf k}_p}\hat{\phi}^{*}_{{\bf k}_p}\rangle
\exp \left ( -\frac{k_p^2}{H}|t_1-t_2| \right ).
\end{equation}
At a given instant, the distribution of 
$\langle \hat{\phi}\hat{\phi}^{*}\rangle$ with respect to the ${\bf k}_p$ 
modes should be the same as that along the ${\bf k}$ modes, since
the relation
between  ${\bf k}_p$ and ${\bf k}$ is specified by a constant multiplier. 
On the other hand, Eqs.(4) and (11) imply that,
$\langle |\phi({\bf k})|^2 \rangle \simeq |\sigma_{\bf k}(\tau)|^2
\propto k^{-3}$ in the super-horizon region. Therefore, 
$\langle \hat{\phi}_{{\bf k}_p}\hat{\phi}^{*}_{{\bf k}_p}\rangle
\propto  k_p^{-3}$, and the integral (20) is infrared divergent. 
However, the duration of the inflation spans only a finite period of time, 
there must be an infrared cutoff at $k \simeq H/R$ with $R\gg 1$. Thus, 
$\langle \hat{\phi}_{{\bf k}_p}\hat{\phi}^{*}_{{\bf k}_p}\rangle$ 
can be approximated as a quantity independent of $k_p$, and Eq.(20) yields
\begin{equation}
\langle \phi({\bf x}, t_1)\phi^{*}({\bf x}, t_2)\rangle \propto 
\left (\frac{t_c}{|t_1-t_2|} \right )^{3/2}, 
   \hspace{3mm} {\rm if} \ \ |t_1-t_2| > t_c,
\end{equation}
where the correlation time $t_c \simeq C/H$, and the constant $C>1$, i.e.  
the correlation time is longer than the Hubble time. Hence, during the
super-Hubble evolution, the field $\phi$ is coherent regardless of the
horizon-crossing moments of the perturbations provided that time
difference is less than $C/H$. This result illuminates that the
gravitational particle creation can be considered as a dissipation source
for the non-equilibrium inflaton $\phi$ during the slow-roll. The
relaxation of that dissipative process is dominated by soft modes and
subsequently gives rise to the classical perturbations with a long-tailed
time correlation. On the other hand, the soft modes do not induce any long
range spatial correlation and therefore, the perturbations in different
spaces ${\bf x}$ retain themselves. That is, the random field of the
inflationary perturbations is ``double-faced'': its spatial distribution
is Gaussian, while the temporal or scaled distribution is coherent. This
property leads to a measurable effect if we use field variables based on a
space-scale decomposition.

Let us consider such space-scale decomposition given by the bases 
$\Psi_{k_p,x}(\bf x')$ which are localized, orthogonal and complete. The 
indices $k_p$ and $x$ denote a cell in the phase ($x$, $k_p$) space, 
i.e. $k_p \rightarrow k_p+\Delta k_p$, $x \rightarrow x + \Delta x$, and 
the differential volume element $\Delta x \Delta k_p \simeq 2\pi$.
Wavelet analysis provides various bases for this sort of space-scale
decomposition\cite{FT}. With a suitable wavelet decomposition, the $\phi$
field can be described by the variables $\delta \phi_{k_p,x}$ defined as 
\begin{equation}
\delta \phi_{k_p,x} = \int \phi({\bf x'}) \Psi_{k_p,x}(\bf x') d{\bf x'}.
\end{equation}
Obviously, $\delta \phi_{k_p,x}$ represents the amplitude of the
perturbation of mode $(k_p,x)$, i.e. the fluctuations of the field at
position around $x$ and on scale around $k_p$.
 
According to Eq.(1), the amplitude $\delta \phi_{k_p,x}$ is mainly
attributed to $\phi({\bf x})$ with the horizon-crossing time corresponding
to $k_p$. Consequently, the time-time correlation of Eq.(21) implies
\begin{equation}
\langle \delta \phi_{k_{p1},x} \delta \phi_{k_{p2},x}\rangle
\neq 0,
\end{equation}
if the difference between the horizon-crossing times corresponding to 
$k_{p1}$ and $k_{p2}$ is less than $C/H$. Equation (23) is a local (at
spatial area around ${\bf x}$) scale-scale ($k_{p1}$ and 
$k_{p2}$) correlation. 

By means of the variables $\delta \phi_{k_p,x}$ one can easily  
perceive this ``double-faces'' feature of the inflationary perturbations.  
The variables $\delta\phi_{k_p,x}$ which are Gaussian with respect to 
$x$ entail  
\begin{equation}
\langle \delta \phi_{k_{p},x}\delta \phi_{k_{p},x'}\rangle = 
P(k_{p})\delta_{{\bf x},{\bf x'}},
\end{equation}
and all higher order cumulants of $\delta \phi_{k_{p},x}$ are zero, 
in which $P(k_{p})$ is the power spectrum of the Gaussian 
fluctuations $\delta \phi_{k_{p},x}$\cite{PF}. Meanwhile, the 
variables at different scales $k_{p}$ may have carried some
correlation. As a simple example, let us consider
\begin{equation}
\delta \phi_{k'_{p},x} = \alpha \delta \phi_{k_{p},x},
\end{equation}
where $\alpha$ is a constant. In this case, the perturbation
$\delta \phi_{k'_{p},x}$ with respect to $x$ is also Gaussian, 
i.e. its variance is $P(k'_{p})=\alpha^2P(k_{p})$ and all higher order 
cumulants of $\delta \phi_{k'_{p},x}$ are zero. However, the 
scale-scale ($k'_p$-$k_p$) correlation at the place is significant, i.e.
$\langle \delta \phi_{k'_{p},x} \delta \phi_{k_{p},x}\rangle
\propto \alpha$. Therefore, a Gaussian power spectrum $P(k_{p})$
can coexist with a local scale-scale correlation [Eq.(23)]. It should 
be emphasized that as a statistical measure, the local scale-scale 
correlation is independent of the Gaussian power spectrum.  
One cannot determine whether there is and/or how strong it is the 
local scale-scale correlation directly by their power spectrum.
  
Recently, the search for the local scale-scale correlation of 
large scale structure  samples has attracted attention.
Using the wavelet technique, the local scale-scale correlations
of QSO Ly$\alpha$ forests\cite{PLGF} and COBE data\cite{PVF}
have been studied. On the other hand, the time-dependent correlation 
of the primordial perturbations is sensitive to the interaction of 
the inflationary field. It is also sensitive to dissipation during 
the inflation. For instance, if the cosmic inflation is thermally 
dissipated\cite{lf}, the correlation time of the initial cosmic 
perturbations will be significantly changed by a thermal-dissipative 
term in Eq.(17). Hence, the local scale-scale correlation behavior of 
the CMB fluctuations and other relevant samples would be useful to 
discriminate among models of inflation.

\acknowledgments

W.-L. L. thanks J.-S. Tsay and S.-Y. Lin for helpful discussion.  He also
acknowledges support from the Republic of China National Science Council
via Grants No. NSC89-2112-M-001-060.

\end{document}